\documentclass{optica-article}

\journal{opticajournal}

\articletype{Research Article}

\usepackage{lineno}
\usepackage{braket}

\usepackage{comment}
\newcommand*\hgzero{\includegraphics[height=7pt]{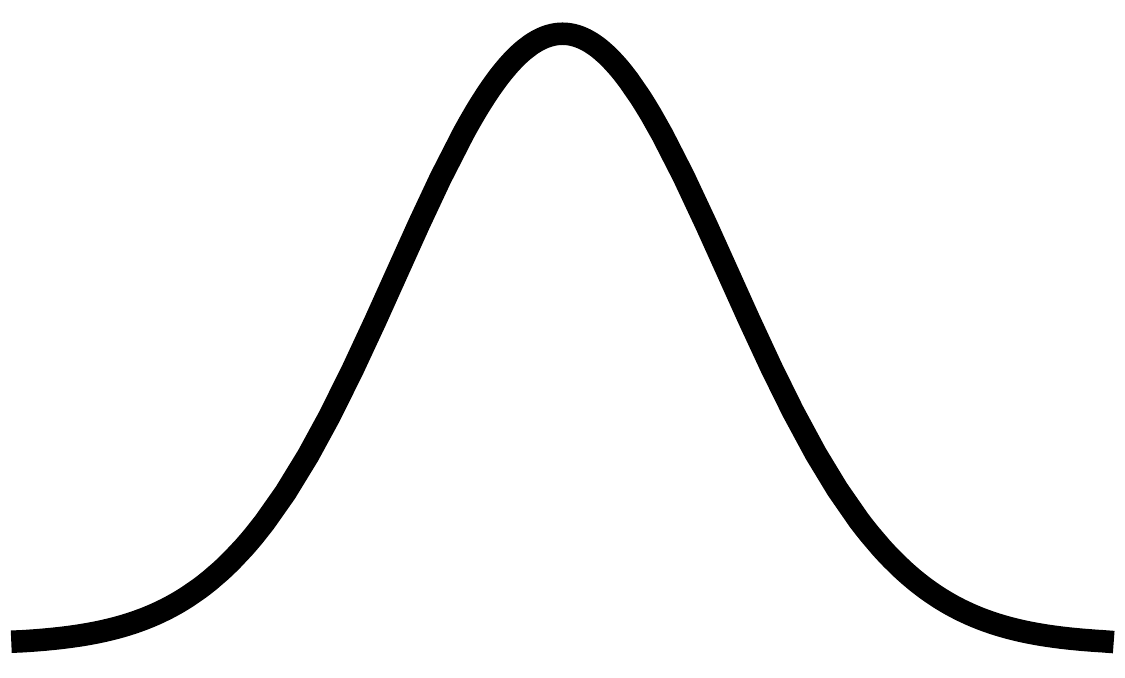}}
\newcommand*\hgone{\includegraphics[height=7pt]{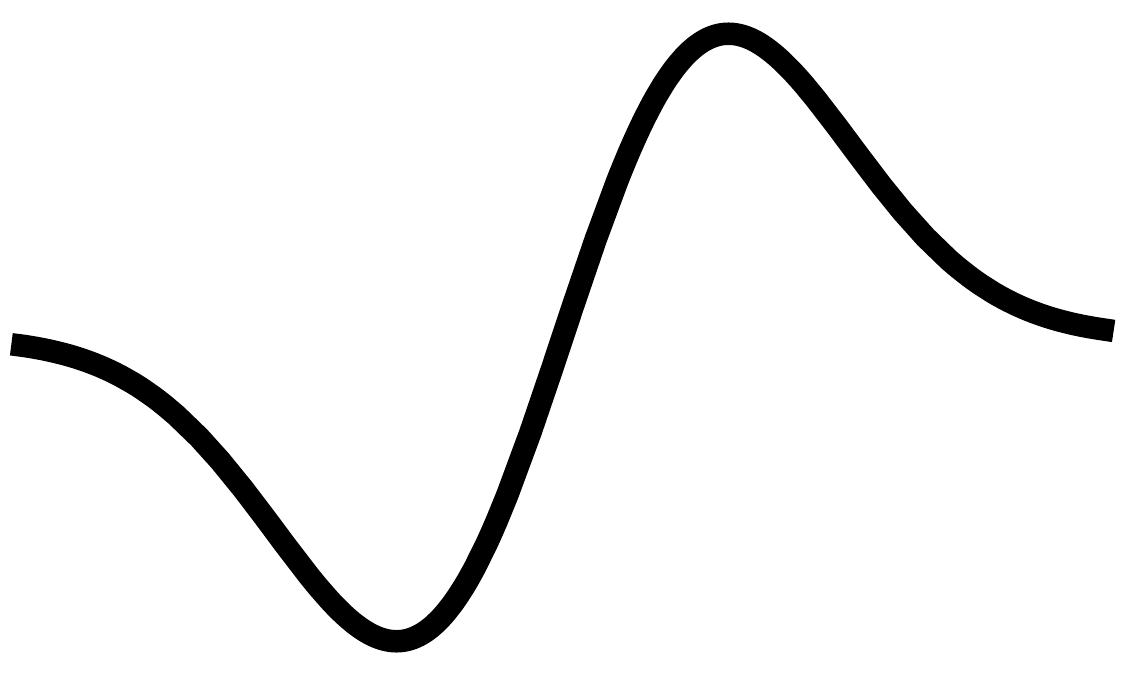}}

\begin{document}

\title{Hyper-entanglement between pulse modes and frequency bins}

\author{Fabrizio Chiriano,\authormark{1,*} Joseph Ho,\authormark{1} Christopher L.~Morrison,\authormark{1} Jonathan W.~Webb,\authormark{1} Alexander Pickston,\authormark{1} Francesco Graffitti,\authormark{1} and Alessandro Fedrizzi\authormark{1}}

\address{\authormark{1}Institute of Photonics and Quantum Sciences, School of Engineering and Physical Sciences, Heriot-Watt University, Edinburgh EH14 4AS, UK}
\email{\authormark{*}fc2005@hw.ac.uk}

\begin{abstract*}
Hyper-entanglement between two or more photonic degrees of freedom (DOF) can enhance and enable new quantum protocols by allowing each DOF to perform the task it is optimally suited for. 
Here we demonstrate the generation of photon pairs hyper-entangled between pulse modes and frequency bins.
The pulse modes are generated via parametric downconversion in a domain-engineered crystal and subsequently entangled to two frequency bins via a spectral mapping technique.
The resulting hyper-entangled state is characterized and verified via measurement of its joint spectral intensity and non-classical two-photon interference patterns from which we infer its spectral phase.
The protocol combines the robustness to loss, intrinsic high dimensionality and compatibility with standard fiber-optic networks of the energy-time DOF with the ability of hyper-entanglement to increase the capacity and efficiency of the quantum channel, already exploited in recent experimental applications in both quantum information and quantum computation.
\end{abstract*}

%%%%%%%%%%%%%%%%%%%%%%%%%%  body  %%%%%%%%%%%%%%%%%%%%%%%%%%
\section{Introduction}
Photons can encode quantum information in multiple degrees of freedom (DOF): polarization, path modes, transverse modes, and the energy-time~\cite{Pan_2012}. 
The energy-time DOF has become increasingly relevant~\cite{Lu_2020,Lu_2022,Reimer_2019,Gil-Lopez_2021, Kues_2017} 
due to its intrinsic multi-dimensional encoding space, which shares with DOF such as 
orbit angular momentum and path mode, its natural resistance to decoherence, its ability to easily generate high visibility interference and compatibility with single mode fibers and waveguides~\cite{Brecht_2015,Lu_2018}.

Within the energy-time domain one can encode quantum information in time bins, frequency bins and pulse modes (which are also known as `temporal' or `time-frequency' modes). 
Both time and frequency bins are particularly relevant in quantum communication and quantum information processing.
They have been shown to be a successful tool to experimentally demonstrate the enhancement between quantum and classical communication protocols~\cite{Wei_2019}, long-distance quantum communication tasks~\cite{Takeda_2013,Kues_2017}, and to decrease the susceptibility to noise and losses and increase key generation rate in quantum key distribution (QKD) protocols~\cite{Nunn_2013,Zhong_2015,Islam_2017}.
Also, recent works have demonstrated new techniques to generate~\cite{Morrison_2022} and characterize~\cite{Gil-Lopez_2021, Lu_2022} discrete bins in the spectral domain, as well as methods to fully manipulate them with a great level of control~\cite{Lu_2018,Lu_2020}.

A pulse mode can be defined as a set of overlapping wave packets in both the time and frequency domain.
A set of pulse modes is identified by their spectral and temporal phases, which preserve the orthogonality between the modes~\cite{Morrison_2022,Brecht_2015}. This natural structure makes them suitable for high-rate transmission, orthogonal-waveform multiplexing, in which signals are distinguished according to the shape of the field~\cite{Reddy_2013}, or to enhance the performance of quantum networks by enabling transmission of photons encoded in different pulse modes using the same communication channel~\cite{Reddy_2014}. 
The direct generation of pulse modes has been demonstrated via parametric downconversion (PDC)~\cite{Graffitti_2020} in crystals with tailored non-linearity profile, while their characterization has been shown using sum frequency generation in waveguides~\cite{Brecht_2011}.
The ability to generate pulse mode entanglement in designed crystals has recently been exploited to also demonstrate entanglement between pulse modes and other high-dimensional DOFs~\cite{Graffitti_vortex}.
This is called \emph{hyper-entanglement}; it allows measurements to be carried out on one DOF without breaking the entanglement structure of the other~\cite{Kwiat_1997,Barreiro_2005}.
The generation of hyper-entangled photons has been demonstrated using multiple sources, such as quantum dots~\cite{Wang_2012}, ion traps~\cite{Hu_2010} and through PDC crystals and waveguides~\cite{Barreiro_2005,Graffitti_vortex,Kang_2014}. 
This scheme is particularly used in quantum communication and quantum information processing to enhance both the channel capacity and efficiency~\cite{Barreiro_2008,Wang_2015,Deng_2017}.
The applications include Bell state analysis~\cite{Kwiat_1998,Williams_2017}, enhanced QKD protocols~\cite{Chapman_2022}, cluster state generation~\cite{Reimer_2019} and the exploration of advanced quantum information concepts in higher-dimensional systems~\cite{Zeitler_2022}.

Here, we report on the generation of a hyper-entangled bi-photon state in pulse mode and frequency bin.
We use a domain-engineered crystal to generate pulse-mode entangled photons through parametric downconversion.
Subsequently, a polarization-frequency bin mapping technique is used to generate the desired hyper-entanglement structure of the bi-photon state.
The characterization of it is performed by reconstructing the joint spectral intensity (JSI), together with the two-photon Hong-Ou-Mandel (HOM) interference measurement~\cite{Hong_1987}.
Furthermore, we perform the Schmidt decomposition on our state to retrieve its energy-time structure and we compare it to the simulated one.

\section{Methods}
In a parametric downconversion process, a pump photon of frequency $\omega_3$ probabilistically generates two photons, traditionally called the signal and idler but here labeled ``1'' ($\omega_1$) and ``2'' ($\omega_2$) respectively, in a non-linear medium.
To a first-order approximation, the bi-photon state can be expressed as,
\begin{equation}
    \ket{\psi} = \iint \,d\omega_1 d\omega_2 \phi(\omega_1,\omega_2) \alpha(\omega_1,\omega_2) \widehat{a}^{\dagger}_1 (\omega_1) \widehat{a}^{\dagger}_2 (\omega_2) \ket{0}_{1,2}.
\label{eq:EPDC_1}
\end{equation}
In the state above, the spectral properties of of the pump field relate to the two downconverted photons via the pump envelope function (PEF), $\alpha(\omega_1,\omega_2)$ which is the result of the energy conservation in the frequency domain.
While often assumed Gaussian, we use a squared hyperbolic secant function for describing the PEF in our simulations as this better matches our mode-locked laser~\cite{Graffitti_2018}.
Additionally, momentum must be conserved during the process which is described by the phase-matching function (PMF) $\phi(\omega_1,\omega_2)$, which depends on the dispersion properties and physical dimensions of the non-linear crystal.
The product of PEF and PMF is the so-called joint spectral amplitude (JSA),
\begin{equation}
    f(\omega_1,\omega_2)=\phi(\omega_1,\omega_2) \alpha(\omega_1,\omega_2),
\end{equation}
which gives a full description of the spectral properties of the bi-photon. 
The JSA can be expressed via the Schmidt decomposition into a suitably chosen set of basis states $g_i(\omega)$ and $h_i(\omega)$, 
\begin{equation}
    f(\omega_1,\omega_2)= \sum_{i=1}^{\infty} \sqrt{\lambda_i} g^{(1)}_i(\omega_1) h^{(2)}_i(\omega_2)
    \label{eq:jsa_shmidt_deco}
\end{equation}
to reveal the inherent energy-time structure of the state~\cite{Ansari_2018}.
Here we choose the Hermite-Gauss polynomial basis for the JSA decomposition.
When $f(\omega_1,\omega_2)$ is normalized, the sum of the coefficients $\lambda_i$ associated with the Schmidt modes is equal to 1 and the Schmidt number $K=1/\sum_{i=1}^{\infty} \lambda^2_i$ returns the number of entangled energy-time modes defining the two-photon state.

\begin{figure}[htb]
\centering
\includegraphics[width=\textwidth]{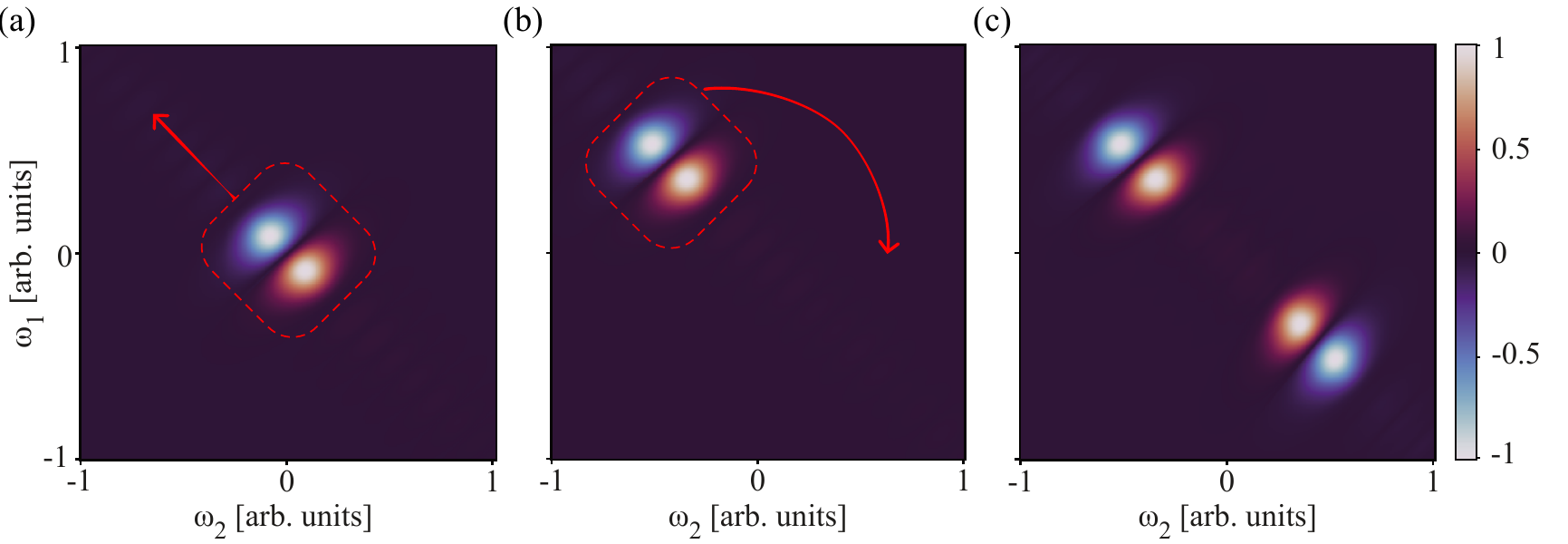}
\caption{\textbf{Process outline in the JSA.}
(a) Starting from the pulse mode entangled state generated by the crystal, the first step is to displace the lobes along the diagonal indicated by the red arrow.
(b) This results in the two downconverted photons being emitted at different wavelengths. The next step is to generate another pair of lobes along the diagonal, noting the lobes are mirrored as indicated by the red arrow.
(c) The final JSA represents the four energy-time modes as described in the main text.
}
\label{fig:fig_1}
\end{figure}

The energy-time structure of the bi-photon state can be manipulated by exploiting the relationship between the PEF~\cite{Reddy_2014,Brecht_2011,Lu_2020} and the PMF~\cite{Graffitti_2017,Graffitti_2020,Pickston_2021,Morrison_2022} in a PDC process.
In our case, we engineer the domain structure of the crystal to tailor its non-linearity as a first-order Hermite-Gauss polynomial.
The resulting bi-photon state is a maximally anti-symmetric Bell state $\vert \psi^{-} \rangle$ encoded in the pulse mode basis $\hgzero = g(\omega)$ and $\hgone = h(\omega)$,
\begin{equation}
    % \ket{\psi} = \frac{\ket{\hgzero{}_1} \ket{\hgone{}_2} - \ket{\hgone{}_1} \ket{\hgzero{}_2}}{\sqrt{2}},
    \ket{\psi} = \frac{\ket{\hgzero{}}_1 \ket{\hgone{}}_2 - \ket{\hgone{}}_1 \ket{\hgzero{}}_2}{\sqrt{2}},
\label{eq:PST_2}
\end{equation}
which generates a JSA characterized by two lobes centered with respect to the photon ``1'' and ``2'' frequency ranges, as show in Figure~\ref{fig:fig_1}(a).

To give the state the sought structure, we first displace it in the frequency space, so that the two daughter photons will be emitted at the different wavelengths.
The effect on the JSA is a shift along the anti-diagonal, representing a decrease in frequency of the one photon and a commensurate increase in frequency of the other (Figure \ref{fig:fig_1}(b)).
The resulting intermediate state is,
\begin{equation}
    % \ket{\psi} = \left( \frac{\ket{\hgzero{}_1} \ket{\hgone{}_2} - \ket{\hgone{}_1} \ket{\hgzero{}_2}}{\sqrt{2}} \right) \otimes \ket{\omega_1} \ket{\omega_2},
   \ket{\psi} = \left( \frac{\ket{\hgzero{}}_1 \ket{\hgone{}}_2 - \ket{\hgone{}}_1 \ket{\hgzero{}}_2}{\sqrt{2}} \right) \otimes \ket{\omega_1}_1 \ket{\omega_2}_2,
\label{eq:PST_3}
\end{equation}
where $\ket{\omega_1}$ and $\ket{\omega_2}$ are two central frequency modes which the two photons are produced.
At this stage, the bi-photon state is entangled only in the pulse mode DOF and fully separable in the central frequencies of the produced photons. 

To obtain entanglement in frequency bins, we first entangle the state in polarization,
\begin{equation}
\ket{\psi} = \left( \frac{\ket{H}_1 \ket{H}_2 - e^{i\phi_p}  \ket{V}_1 \ket{V}_2}{\sqrt{2}} \right)  \otimes 
 \left(\frac{\ket{\hgzero{}}_1 \ket{\hgone{}}_2 - \ket{\hgone{}}_1 \ket{\hgzero{}}_2}{\sqrt{2}} \right) \otimes 
 \ket{\omega_1}_1 \ket{\omega_2}_2,
\label{eq:EPDC_2}
\end{equation}
where $\phi_p$ is the phase associated to the polarization DOF.
This intermediate step allows us to map the polarization DOF to the spectral domain~\cite{Ramelow_2009}, and once we then erase the polarisation information in state Eq.~\eqref{eq:EPDC_2} we finally recover the desired energy-time structure, as in Figure~\ref{fig:fig_1}(c). 
The state produced has a phase $\phi_p$ associated with the polarization domain which can be manipulated using phase retarders. 
The final state can be expressed as the tensor product of two pulse modes and two frequency bins:
\begin{equation}
\ket{\psi} = \left( \frac{\ket{\omega_1}_1 \ket{\omega_2}_2 - e^{i\phi_p}  \ket{\omega_2}_1 \ket{\omega_1}_2}{\sqrt{2}} \right)  \otimes \left(\frac{\ket{\hgzero{}}_1 \ket{\hgone{}}_2 - \ket{\hgone{}}_1 \ket{\hgzero{}}_2}{\sqrt{2}} \right).
\label{eq:EPDC_3}
\end{equation}

\begin{figure}[htb]
\centering
\includegraphics[width=\textwidth]{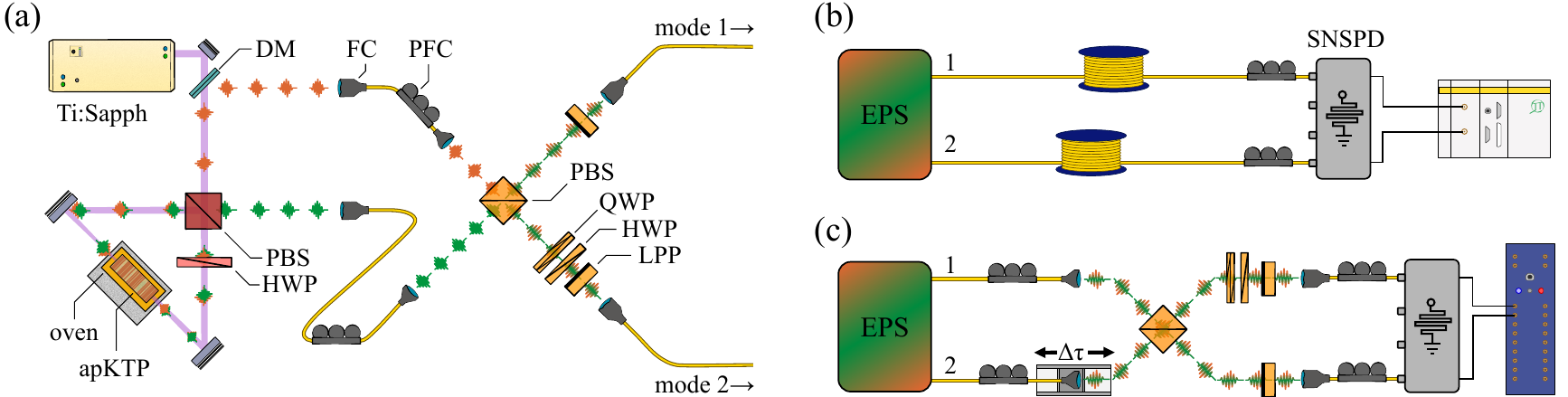}
\caption{\textbf{Experimental layout.}
(a) Entangled photon source (EPS). A Ti-Sapphire laser pumps the aperiodically-poled KTP (aKTP) crystal mounted in a temperature controlled oven.
The crystal in embedded in a Sagnac loop using a dual-coated polarizing beamsplitter (PBS) and dual-coated half-wave plate (HWP).
Both downconverted photons are sent to single-mode fiber couplers (FC), in one arm a dichroic mirror (DM) is used.
The polarization-to-frequency mapping setup consists of polarization fiber controllers (PFC), PBS, quarter-wave plate (QWP), HWP and linear plate polarizers (LPP).
(b) JSI measurement. The time-of-flight spectrometer is realized by adding 20~km spooled SMF28 fiber to each arm of the EPS.
Photons are detected by superconducting nanowire single photon detectors (SNSPD) and arrival times are timetagged by a fast logic unit (PicoQuant).
(c) Two-photon interference measurement.
The two-photon interferometer consists of PFCs, PBS, QWP, HWP and LPPs.
Additionally, one photon from the EPS is temporally varied, $\tau$, using a motorized stage.
Detected photons are recorded using a multi-channel logic unit (UQDevices).
}
\label{fig:fig_2}
\end{figure}

The experimental setup for generating the frequency bin and pulse mode entangled photon pairs is shown in Figure~\ref{fig:fig_2}~(a). 
We use a mode-locked Ti:Sapphire laser with a $80$~MHz repetition rate, 1.27~ps pulse duration and central wavelength of 775~nm to pump a collinear type-II PDC process.
The pump laser is focused into a 30~mm long aperiodically poled potassium titanyl phosphate (apKTP) crystal, whose PMF approximates first order Hermite-Gauss function shape using our coherence length domain algorithm~\cite{Graffitti_2020}.
We use a custom oven to heat the crystal to $\sim 135^\circ$C to shift the downconverted photons in frequency such that the final bin separation is expected to be $\sim11$~nm.
This is done to ensure minimal overlap between the frequency bins, which have a nominal bin width of $\sim3$~nm as measured in Ref.~\cite{Graffitti_2020}.
At this step, we obtain the bi-photon state given in Eq.~\eqref{eq:PST_3}.

The crystal is embedded in a Sagnac interferometer to entangle the photon pairs in polarization~\cite{Fedrizzi_2007}, creating hyper-entanglement in the polarization and pulse mode DOF as in Eq.~\eqref{eq:EPDC_2}.
Subsequently, the two photon are sent through two single-mode fibers (SMF) and impinge on a polarizing beam splitter (PBS). 
In one of the output mode of the PBS, we set up a quarter-wave plate (QWP), half-wave plate (HWP) and linear plate polarizer (LPP) to control the polarization phase $\phi_p$, while in the other output mode we use a single LPP.
Both LPPs are set to project the photons onto the diagonal/anti-diagonal ($\ket{D},\ket{A}$) basis, which together with the PBS makes them indistinguishable in polarization, thus translating the polarization entanglement to spectral entanglement.
The state produced by the source (labeled EPS in Figure~\ref{fig:fig_2}) has the sought hyper-entangled structure in frequency bin and pulse mode, as in Eq.~\eqref{eq:EPDC_3}.

This polarisation mapping technique has a success probability of $25\%$, due to the use of the PBS and the LPP for the polarization erasure operation, both of which incur a 50\% loss.
The overall measured brightness of this source of hyper-entanglement is $\sim 0.6$ kHz/mW.
This matches with expectations given the brightness measured for the pulse mode crystal in~\cite{Graffitti_2020} subject to the success probability of the mapping technique and optical losses in this setup. 
The photons are detected using commercial superconducting nanowire single photon detectors (SNSPDs).

\section{Results}
There are several approaches that can ideally allow complete characterization of the energy-time DOF of the target bi-photon state.
The use of quantum pulse gates~\cite{Brecht_2011} and quantum frequency processors~\cite{Lu_2020} would enable complete tomography of the state in the frequency and temporal domains, respectively.
In addition, compressive tomography techniques, which exploits precise knowledge of the system state space, can be applied to both DOF~\cite{Gil-Lopez_2021}.
Other less direct methods use cross-measurement in time-frequency space so that spectral and temporal phases can be reconstructed by measuring JSI and JTI~\cite{MacLean_2019}.

In our experiment, we reconstruct the joint spectral intensity of the state via time-of-flight spectroscopy using fast counting logic for processing detection events.
Additionally, we perform a HOM interference measurement, reconstructing the bi-photon interferogram in time, using a multichannel counting logic. 
We exploit \emph{a priori} knowledge of the desired energy-time entanglement structure to combine the results of the JSI and the HOM interference patterns to reconstruct the JSA. 
Finally, we show that spectral phase in the biphoton state results in a unique set of real and imaginary JSA, allowing us to identify the the reconstructed JSA. 

\subsection{JSI measurement}
\begin{figure}[htb]
\centering
\includegraphics[width=1\textwidth]{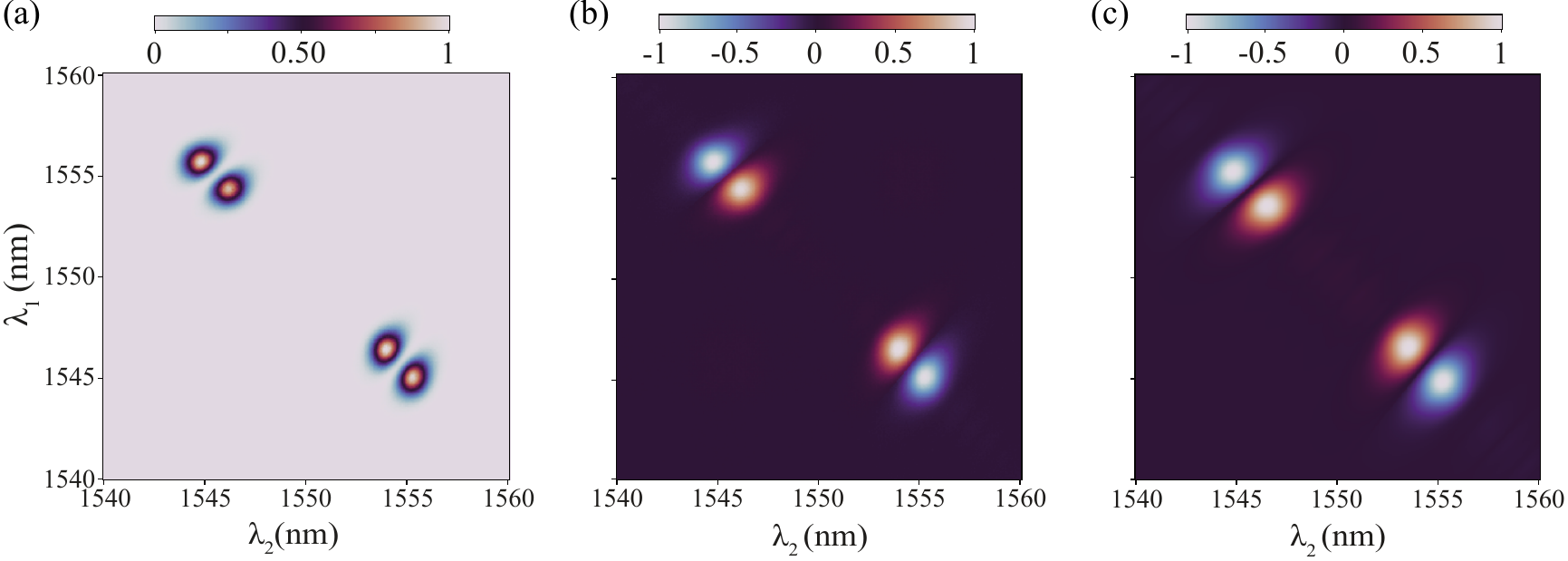}
\caption{\textbf{JSA reconstruction.} (a) Measured JSI in a 20~nm range, consisting of $320\times 320$ bins with 0.0625~nm width.
(b) We apply a $\pi$ phase profile to the measured JSI to infer the JSA of the measured state.
(c) The simulated JSA for the ideal frequency bin and pulse mode hyper-entangled state for comparison.}
\label{fig:fig_3}
\end{figure}

Our time-of-flight spectrometer (TOFS) is realized by adding dispersive elements to each photon after the source as shown in Figure~\ref{fig:fig_2}(b) and measuring the arrival times of each photon.
We use two $\sim$20~km single-mode (SMF28) fibers which have a nominal chromatic dispersion of $\sim$20 ps/nm/km at 1550~nm, however the fibers also introduce $\gtrsim8$~dB loss to each photon.
To collect enough statistics for reconstructing the JSI we use measurement integration times of approximately twelve hours to obtain  $>1.3 \times 10^{7}$ total coincidences.
The coincidence events are established from time-tagged single photon detection events recorded using a \emph{Hydroharp 400} which has a $1$~ps resolution.
We can then reconstruct the JSI over a $36$~nm spectral range, corresponding to $12.5$~ns which is the pulse separation of the Ti-Sapph laser, with a spectral bin width of 0.0625~nm, corresponding to $25$~ps.
In Figure~\ref{fig:fig_3}(a), the JSI is shown for a $20$~nm range which contains the main features of the bi-photon spectra.
The SNSPD system has a nominal quantum efficiency of $\sim$80$\%$ and temporal jitter of $<$50~ps which ultimately sets the minimal spectral resolution of our TOFS.

As shown in Figure \ref{fig:fig_3}, the measured JSI exhibits the same behavior as the simulations, respectively panel (a) and (c), displaying two distinct pairs of lobes. 
The presence of two lobes in each pair is indicative of the pulse mode entanglement, while the separation of the pairs is a consequence of the frequency entanglement.

To verify the quantum state in terms of energy-time modes, we perform the Schmidt decomposition on the reconstructed JSA and calculate the Schmidt number $K$, with $K>1$ to indicate entanglement. 
We first calculate the expected Schmidt number by simulating the effects of applying heating to the crystal to induce the non-degeneracy of the photon pair emissions.
Primarily, this leads to expansion of the crystal, thus the poling domain widths, and also modifying the refractive indices according to the Sellmeier equation. 

From our theoretical model for the JSA, we expect a Schmidt number $K_{theor} = 4.02$. 
To compare this with our experimental results we need to infer a JSA from the measured JSI data, which lacks spectral phase information.
Based on the results in Ref.~\cite{Graffitti_2020}, we can state the existence of a $\pi$ phase between the two lobes of each pulse-mode pair. This phase is associated with the first order Hermite-Gauss function, which is the shape designed in the crystal non-linearity using our domain engineering technique. 
In addition, we will assume that the frequency bin entanglement has been prepared with a globally symmetric structure, which applies an additional $\pi$ phase difference between the two frequency bins, give the anti-symmetry of the state in the pulse mode DOF. We will verify that this choice is the only one consistent with the observed two-photon interference signature in the next section.
We can now apply this global phase profile and derive the JSA of the state (Figure \ref{fig:fig_3}~(b)). We perform a Schmidt decomposition on the reconstructed JSA and obtain a Schmidt number of $K_{\sqrt{\text{JSI}}+\text{phase}}=4.0391 \pm 0.0003$. This uncertainty corresponds to $3\sigma$ standard deviations, obtained from $10^3$ rounds of Monte Carlo simulation, which implies excellent agreement with the expectation based on the energy-time structure of the state in \eqref{eq:EPDC_3}.

\subsection{Hong-Ou-Mandel interference measurement}
In this section we analyze the measurement of state symmetry information, from which we reconstruct the phase information. 
This is done by performing an HOM style interference measurement and reconstructing the bi-photon interferograms for a set of phase values.
From the different HOM patterns we can then conclude coherent interference between the two photons and, furthermore, infer the global symmetry of the bi-photon state.

To measure the HOM patterns we remove the 20~km fibers used for the JSI measurement and send the two photons from the hyper-entanglement source to an interferometer.
More specifically, the two photons interfere in a PBS, thus requiring polarizers to project in non-polarization discriminating fashion to correctly obtain the HOM effect, and we measure the two-fold coincidences for a range of time delays, $\tau$ (Figure \ref{fig:fig_2}~(d)).
The coincidence probability can be analytically expressed as below, and it is used to fit the data points from the HOM measurements:
\begin{equation}
\begin{split}
p_{cc}= &N \left(\frac{1}{2} - \frac{V}{8} e^{-i\delta\tau - \frac{\sigma^2\tau^2}{4} - i\phi} \times \right. \\
&\left. \frac{ e^{\frac{\delta^2}{\sigma^2}}(1+e^{2i(\delta\tau+\phi)})\sigma^2(\sigma^2\tau^2-2) + 2 e^{i(\delta\tau+\phi)}(4\delta^2 - 2\sigma^2 + \sigma^4\tau^2) 
}{e^{\frac{\delta^2}{\sigma^2}}\sigma^2 - 2\delta^2\cos{\phi}+\sigma^2\cos^2{\phi}}\right)
\end{split}
\label{hom_expression}
\end{equation}
Here, $\tau$ is the time delay between the two photons, $\delta$ is the frequency bin separation, $2\sigma$ is the width frequency bin, $\phi$ is the global phase of the state, $V$ is the visibility of the interferogram, and $N$ denotes the normalized counts.

\begin{figure}[htb]
\centering
\includegraphics[width=\textwidth]{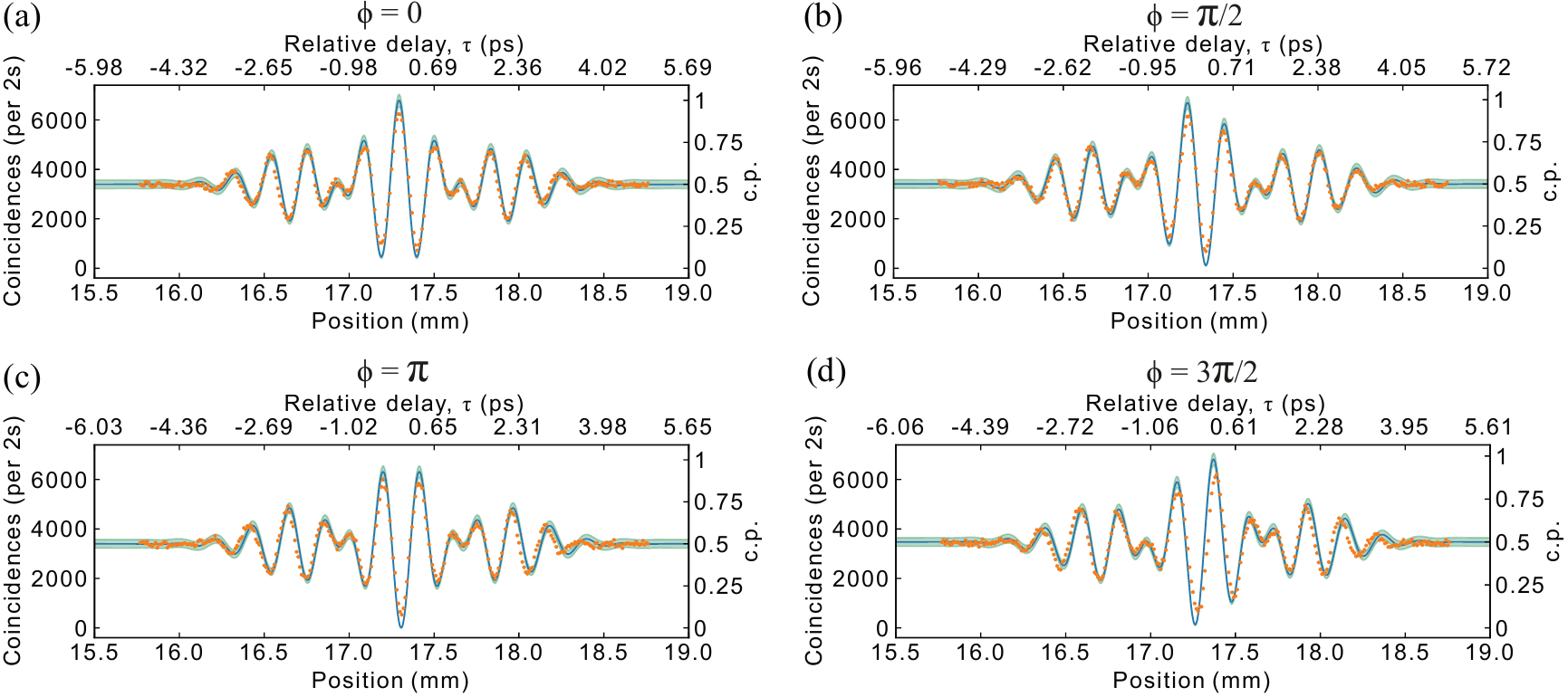}
\caption{\textbf{Two-photon interference patterns.}
The measured interferograms corresponding to different $\phi$ are shown in plots (a) - (d), with the respective phases denoted above.
Raw coincidence data (orange dots) are plotted as a function of delay stage position (bottom axis) alongside the corresponding temporal delay, $\tau$ (top axis).
Theoretical curves (blue line) based on ideal simulations are scaled to the normalized coincidence probability (c.p.) displayed on right axis.
A three-sigma confidence region (green) is shown, assuming Poissonian statistics.
} 
\label{fig:fig_4}
\end{figure}

Figure~\ref{fig:fig_4} shows the plots for the experimentally measured HOM interferograms for four different phase values, alongside the simulated theory curves.
For clarity, the fitted interferogram is not shown.
For the four nominal phases prepared, $\phi= \{ 0,\pi /2, \pi,3\pi/2 \}$ we obtain fits for each interferogram and observe the measured phases to be $\{ 0.031\pm 0.086, 1.61\pm0.07, 3.19\pm0.1, 4.61 \pm 0.16 \}$ rad.
Notably, for $0$ and $\pi$ phases, the measured interferograms clearly reproduce the typical anti-bunching behaviour for the antisymmetrical state and the bunching for the symmetrical case.
We include the results for the two intermediate phase cases, $\phi=\{pi/2,3\pi/2\}$, which highlights our ability to precisely control the phase of the bi-photon state.
In all cases we observe good agreement between the measured phase and the nominal phase setting.

In the case of perfect two-photon interference, the expected visibility is unity.
For the four cases in our experiment the fitted visibility values were $\{0.863\pm0.012, 0.844\pm0.011, 0.841\pm0.014, 0.819\pm0.022\}$.
The reduction in visibility is primarily attributed imperfect mode-matching which leads to partial distinguishability between the two photons thus degrading the measured visibility.
In our two-photon interferometer the erasure of the polarization information after the PBS plays a key role, however the wavelength dependence in the polarization optics leads to slightly different waveplate transformations corresponding to each frequency bins.
Moreover, as the central wavelength of the frequency bins are shifted away from 1550~nm, this introduces added loss when transmitting through optical devices and fibers.

We use the fitted interferograms to also estimate the spectral separation as well as the spectral bandwidths of the two frequency bins of the bi-photon state.
These are captured by the parameters, $\delta$ and $2\sigma$ respectively in \eqref{hom_expression}.
Notably, these values correspond to the angular frequencies, thus requires division by $2\pi$.
From the fit we obtain a bin separation of $1.37 \pm 0.4$ THz, while the bin width is $0.38 \pm 0.06$ THz, which corresponds to $\sim11$~nm separation and $\sim3$~nm bin width, assuming the central wavelength at 1550~nm.

For all the measured HOM scans, the overlap of theoretical and measured HOM interference patterns is well within the $3\sigma$ confidence region, indicating our ability to accurately control the bi-photon state. Errors on the parameters are quoted as standard errors from the fit.

\subsection{Inferring phase information from bi-photon interferogram}
The JSA provides a complete description of the bi-photon state in the energy-time DOF. Measuring both the JSI and phase information would allow reconstructing the former and fully characterising the state. From a mathematical perspective, having an amplitude and a phase is equivalent to representing the JSA as a complex function, so that it can be decomposed into its real and imaginary parts. Although conventionally its representation is confined to the real part, this may not be sufficient to characterize the state entirely, as we will show here.

\begin{figure}[htb]
\centering
\includegraphics[width=\textwidth]{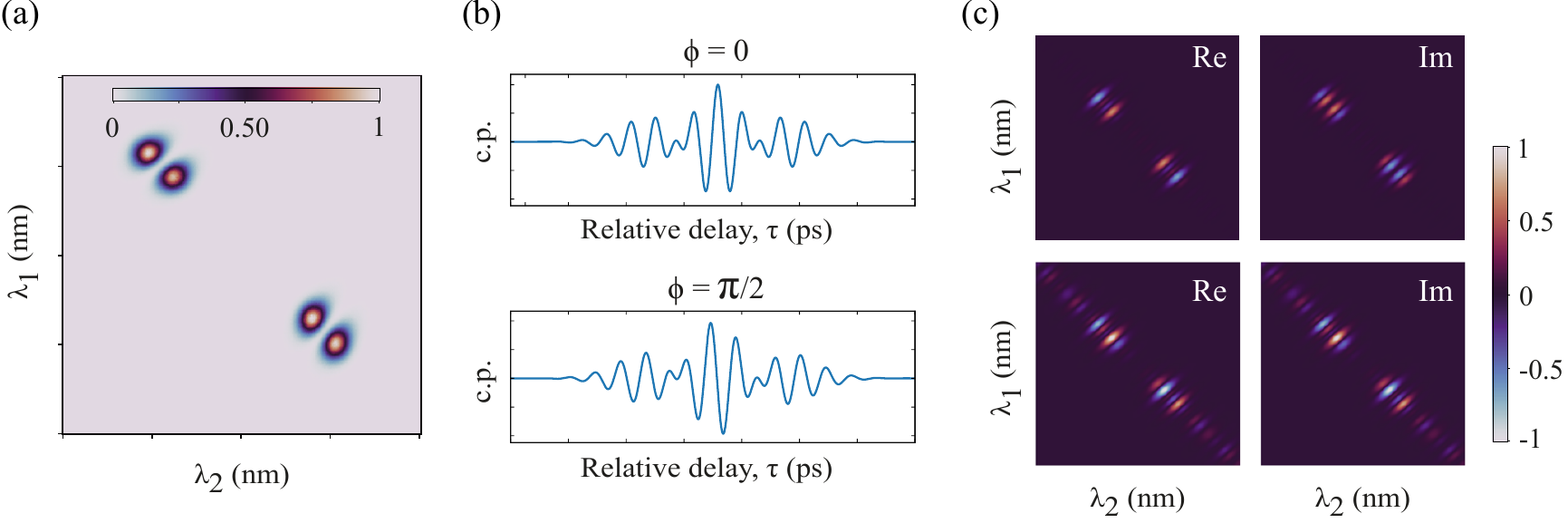}
\caption{\textbf{JSA derivation through decomposition.} (a) Experimentally measured JSI. (b) Simulated bi-photon interferogram for $\phi = 0$ and $\phi = \pi /2$ phase settings. (c) By simulating the design of two crystals that would emit the state with the two phase values, we can produce the JSAs that would be experimentally measured through direct phase measurements. The figure shows the JSA decompositions in their real and imaginary components.} 
\label{fig:fig_5}
\end{figure}

Figure~\ref{fig:fig_5} presents the procedure by which the measured state can be unambiguously identified.
Figure~\ref{fig:fig_5}~(a) shows the JSI measured in the laboratory, while Figure~\ref{fig:fig_5}~(b) shows the simulated HOM patterns for two phase values.
The JSI does not give us any phase information, which means that for different phase values the measured JSI is identical.
On the other hand, the HOM interferogram indirectly gives us information about the phase of the state by revealing its the global symmetry. Still, it is not sufficient on its own to deduce the entire structure of the state.
This ambiguity due to the lack of information does not immediately allow us to assert what the JSA of the analysed state is. However, it can be circumvented.
Considering the phase values $\phi=0$ and $\phi=\pi/2$ recoverd by the HOM scans, we design through simulations the crystals that would actually emit the state associated to these phases, and we obtain the two correspondent JSAs that the two crystals would produce.
It is evident from Figure~\ref{fig:fig_5}(c) that the two JSAs differ from their decomposition into real part and imaginary part, which is alway unique, thus resulting in JSAs proper to two different states.
This gives us strong evidence that the phase obtained from the HOM scans can be used to unambiguously identifies the analysed state.
We can, therefore, state that the inferred JSA shown in the Figure~\ref{fig:fig_3}(b) can be unambiguously deduced even in the absence of direct measurements on the phase of the state. 

\section{Discussion}
We have experimentally implemented and characterized a photon pair source that is hyper-entangled, in frequency bins and pulse mode, using a polarization-mapping technique~\cite{Ramelow_2009} to create the frequency bin entanglement.
In a recent experiment in Ref.~\cite{Ruibo_2021}, an alternative method was introduced for mirroring a joint bi-photon spectrum around a central frequency, using a PDC crystal in a linear double-pass configuration.
One advantage of that method is that it does not suffer from the 50$\%$ loss which is incurred in the entanglement transfer from polarization to frequency.
However, the use of the polarization-mapping allowed for precise phase control, by employing waveplates to encode the phase in polarization which is then transferred to the spectral domain.

The intrinsic structure of our hyper-entangled state could be exploited in fundamental tests of quantum mechanics or multiphoton quantum metrology applications~\cite{Lyons_2018}.
A possible application is the all-optical analogous Stern-Gerlach effect.
This has been recently demonstrated through sum frequency generation processes to show the analogy between the spin of the atom and the frequency of light~\cite{Yesharim_2022}. 
Using our state, this effect could be shown through PDC process and the hyper-entangled structure could give rise to effects not studied so far.

We used time-of-flight spectroscopy and two-photon interference to indirectly infer the hyper-entanglement structure of the biphoton state.
The logical next step would be to verify the independence of frequency bin and pulse mode DOFs respectively by tracing out one and confirming entanglement is preserved in the other.
For example, one can use quantum pulse gates (QPG) to project each photon in a specific pulse mode~\cite{Ansari_2018,Serino_2022} to preserve the frequency bin structure which can then be completely characterized using quantum frequency processor (QFP)~\cite{Lu_2020}.
Similarly using QFPs, or simply diffraction gratings, first to sort each photon into their frequency bin followed by characterization of the pulse-mode entanglement structure using QPGs\cite{Serino_2022}.
The structure of hyper-entanglement in frequency bin and pulse mode may be exploited in quantum communication networks that use wavelength de-multiplexing (WDM) devices to distribute pulse mode entangled qubits to pairs of users~\cite{Soren_2018}.
By encoding in pulse modes, one can take advantage of an increased quantum channel capacity by invoking a larger alphabet, as well as improved robustness to noise in transmission through fibers.

The feasibility of these applications will require photons with narrower bandwidth for compatibility with devices used in the QFP and QPG, as well as commercial WDMs in existing networks.
This may be overcome by applying spectral compression techniques on each photon using chirped upconversion \cite{lavoie2013spectral} or time lenses~\cite{karpinski2017bandwidth}.
Alternatively, it remains an open challenge to design a tailored non-linear crystal that can simultaneously achieve narrow bandwidth photons and produce hyper-entanglement across multiple energy-time DOFs.

\section{Acknowledgements} 
We thank P. Barrow for his technical assistance in the TOFS measurements.

\section{Funding} This work was supported by the UK Engineering and Physical Sciences Research Council (Grant Nos, EP/T001011/1.).
F G acknowledges studentship funding from EPSRC under Grant No. EP/L015110/1.

\section{Disclosures} The authors declare no competing interest.

\section{Data availability} Data underlying the results presented in this paper are not publicly available at this time but may be obtained from the authors upon reasonable request.

% \bibliography{sample}

\end{document}